\documentclass[aps,twocolumn,preprintnumbers]{revtex4}

\usepackage{graphicx}  
\usepackage{subfigure}
\usepackage{multirow}
\usepackage{tabularx}
\usepackage{fancyhdr}
\usepackage{longtable}
\usepackage{parskip}
\usepackage[T1]{fontenc}
\usepackage{dcolumn}   

\usepackage{bm}        
\usepackage{amsfonts}  
\usepackage{amsmath}   
\usepackage{amssymb}   

\begin{document}

\title{Hyperscaling analysis of a disorder-induced ferromagnetic quantum critical point in Ni$_{1-x}$Rh$_{x}$ with $x = 0.375$}

\author{R.-Z. Lin}

\affiliation {Department of Physics, National Cheng Kung University, Tainan 701, Taiwan}
\affiliation {Center for Quantum Frontiers of Research \& Technology (QFort), National Cheng Kung University, Tainan 701, Taiwan}

\author{C.-L. Huang}

\affiliation {Department of Physics, National Cheng Kung University, Tainan 701, Taiwan}
\affiliation {Center for Quantum Frontiers of Research \& Technology (QFort), National Cheng Kung University, Tainan 701, Taiwan}

\date{\today}

\begin{abstract}  

Here we report on hyperscaling analysis on thermodynamic measurements as a function of temperature and magnetic field for Ni$_{1-x}$Rh$_{x}$ with $x = 0.375$ where a ferromagnetic quantum critical point has been recently identified [Phys. Rev. Lett. $\textbf{124}$, 117203 (2020)]. The obtained critical exponents agree well with the theory proposed by Belitz, Kirkpatrick, and Vojta for disorder tuned quantum critical point in the preasymptotic region. 

\end{abstract}


\maketitle 

\section{Introduction}

A magnetic quantum critical point (QCP) occurs when a second-order phase transition is suppressed to absolute zero by applying a nonthermal control parameter, such as, pressure, magnetic field, strain, or chemical substitution \cite{Loehneysen2007}. A good deal of strange phenomena have been discovered in the vicinity of the QCP. These phenomena are often accompanied by a breakdown of Fermi-liquid (FL) theory, and in some cases, unconventional superconductivity \cite{schuberth2016emergence,Ran2019}. It is still unclear whether the quantum fluctuation of order parameter, the only relevant energy scale close to the QCP, intricately connects the quantum criticality and unconventional superconductivity \cite{Nakai2010,Keimer2015,Ran2019}. Nevertheless, exploring new QCPs could potentially pave the way for deeper insight in understanding of the non-Fermi liquid physics and shed light on the relation between the quantum fluctuations and unconventional superconductivity. 

Successful tuning of a magnetic system towards the QCP has proven rarer in ferromagnets than in antiferromagnets, most probably due to the existence of multiple dynamical exponents in the former \cite{Manuel2016}. Although QCPs have been revealed in ferromagnets Zr$_{1-x}$Nb$_{x}$Zn$_{2}$, SrCo$_{2}$(Ge$_{1-x}$P$_{x}$)$_{2}$, YbNi$_{4}$(P$_{1-x}$As$_{x}$)$_{2}$, and CeRh$_{6}$Ge$_{4}$ \cite{Sokolov2006,Jia2011,Steppke2013,Shen2020}, a decisive route to reach the ferromagnetic (FM) QCP remains ambiguous. A recent theory proposed by Belitz, Kirkpatrick, and Vojta (BKV) suggests one may restore the FM QCP by introducing appropriate amount of quenched disorder \cite{Belitz1999PRL,Sang2014,Kirkpatrick2014,Kirkpatrick2015}. To look for model systems to test the BKV theory, one shall start from the one with a simple crystal structure to avoid structure complexity. In addition, the $f$-electron FM system is less appropriate compared to the $d$-electron ones, as the disorder also adjusts the competition between magnetic RKKY interactions and nonmagnetic Kondo effect in the $f$-electron magnets \cite{Stewart2001,Stewart2006}. Ni has a simple face-centered cubic structure and is a three-dimensional $d$-electron ferromagnet with a Curie temperature $T_{C} \approx 627$ K \cite{Howard1965,Kraftmakher1997}. The ordering temperature can be suppressed by alloying Ni with nonmagnetic or paramagnetic transition metals Ni$_{1-x}A_{x}$
\cite{Gupta1964,Brinkman1968,Boelling1968,Gregory1975}. The low-temperature quantum critical behavior has been extensively studied near the critical concentration $x_{c}$ in Ni$_{1-x}$Pd$_{x}$ \cite{Nicklas1999}, Ni$_{1-x}$V$_{x}$ \cite{Ubaid-Kassis2010}, and Ni$_{1-x}$Rh$_{x}$ \cite{Huang2020}. The former two alloys did not exhibit QCPs, but showed randomness in magnetic interactions caused by disorder: superparamagnetism was suggested to be responsible for constant thermal expansion coefficient as $T \rightarrow 0$ in Ni$_{1-x}$Pd$_{x}$ \cite{Kuechler2006}, and additional disorder-induced fluctuations gave rise to a nonanalytic contribution to the free energy, forming so called the Griffiths rare regions, in Ni$_{1-x}$V$_{x}$ \cite{Ubaid-Kassis2010}. Our previous work on Ni$_{1-x}$Rh$_{x}$ presented thermodynamic evidence for a disorder-induced FM QCP in the vicinity of $x_{c} = 0.375$ (see Fig.~\ref{Fig1}(a)) \cite{Huang2020}. To the best of our knowledge, Ni$_{1-x}$Rh$_{x}$ is the only Ni$_{1-x}A_{x}$ alloy that exhibits a FM QCP. Moreover, it shows the first occurrence of the FM QCP with direct dilution of the $d-$electron magnetic site, leading Ni$_{1-x}$Rh$_{x}$ to held a unique position among quantum critical materials. 

Close to a QCP universal relations among thermodynamic properties appear. That is, a set of critical exponents and the scaling functions are universal up to a certain symmetry and spatial dimensionality \cite{Hohenberg1977}. Here we report hyperscaling analysis of the low-temperature magnetization and specific heat data for Ni$_{1-x}$Rh$_{x}$ with $x = 0.375$, bearing all the features of a FM QCP \cite{Huang2020}. The obtained scaling exponents are in line with the theoretical prediction for a FM fixed point in the asymptotic limit of high disorder \cite{Kirkpatrick2014,Kirkpatrick2015}. This work reinforces the role of quenched disorder in tuning the FM transition continuously to absolute zero, as suggested by the BKV theory, providing a promising route to enlarge the portfolio of FM quantum critical materials, in addition to antiferromagnetic counterparts.

\section{Experimental details}
Polycrystalline Ni$_{1-x}$Rh$_{x}$ samples were prepared by arcmelting high-purity Ni and Rh elements. The details of sample characterization had been reported in our earlier work \cite{Huang2020}. Isothermal magnetization measurements were carried out between the temperature $T =$ 1.8 and 20 K and the magnetic field $B =$ 0 and 7 T using a Quantum Design (QD) magnetic property measurement system. Samples were field cooled at 7~T from high temperatures before performing the measurements. Specific heat was measured between $T =$ 0.05 and 30 K and $\mu_{0}H =$ 0 and 14 T using a QD physical property measurement system equipped with  a dilution refrigerator option.

\section{Results}
\begin{figure}
\includegraphics[width=1.1\columnwidth]{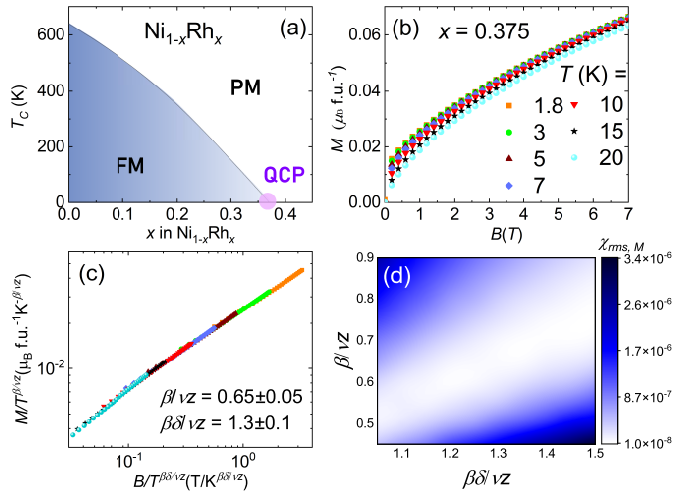}
\caption{(a) $T_{C}-x$ phase diagram of Ni$_{1-x}$Rh$_{x}$. FM stands for ferromagnetically ordered state and PM stands for paramagnetically ordered state. The phase boundary between FM and PM is from Ref.~\cite{Huang2020} and references therein.(b) Isothermal magnetization $M$ curves for the $x = 0.375$ sample.(c) Scaling of $M$ as a function of $T$ and $B$.(d) Mean square deviation $\chi_{rms,M}$ as a function of critical exponents $\beta/\nu z$ and $\beta\delta/\nu z$.}  
\label{Fig1} 
\end{figure}

The magnetic isotherms of Ni$_{1-x}$Rh$_{x}$ with $x = 0.375$ are shown in Fig.~\ref{Fig1}(b). $M(T,B)$ increases less rapidly with increasing $T$, suggesting the applicability of a scaling relation. At a QCP below the upper critical dimension, the scaling relation of the Gibbs free energy $\mathcal{F}(T,B)$ reads
\begin{equation}
    \mathcal{F}(T,B)=b^{-(d+z)}\mathcal{F}\left ( b^{z}T,b^{\beta\delta/\nu}B\right ),
\end{equation}
where $b$ is an arbitrary scale factor, $z$ is a dynamical exponent associated with the tuning parameter $T$, and $\beta\delta/\nu$ is the scaling exponent associated with the tuning parameter $B$ \cite{Loehneysen2007}. Customarily, $\beta$ describes the concentration dependence of the order parameter $m(r=x_{c}-x,T=B=0)\propto r^{\beta}$, $\delta$ narrates the field dependence of $m (r=T=0,B) \propto B^{1/\delta}$, and $\nu$ is the exponent obtained from the scaling relation of diverging correlation length $\xi \propto |r|^{-\nu}$. For a FM system like Ni$_{1-x}$Rh$_{x}$, the order parameter is the magnetization $M$. The field and temperature dependence of $M$ in the vicinity of a QCP follows:
\begin{equation}
    M(T,B)=-\frac{\partial\mathcal{F}}{\partial B}=b^{\beta\delta/\nu-(d+z)}M\left ( b^{z}T,b^{\beta\delta/\nu}B\right ).
\end{equation}
If we choose $b^{z}T = T_{0}$ with the cutoff energy $k_{\rm B}T_{0}$, one gets
\begin{equation}
    \frac{M(T,B)}{T^{\beta/\nu z}}=\Phi \left( \frac{B}{T^{\beta\delta/\nu z}} \right ),
\end{equation}
indicating that $M/T^{\beta/\nu z}$ should be a universal function of $B/T^{\beta\delta/\nu z}$. Excellent scaling over three orders of magnitude of $B/T^{\beta\delta/\nu z}$ with $\beta/\nu z = 0.65\pm0.05$ and $\beta\delta/\nu z = 1.3\pm0.1$ is shown in Figure~\ref{Fig1}(c). Due to domain effects, the data of low fields ($B < 1$ T) 
are omitted \cite{Huang2015,Huang2016}. The trend of scaling plot is similar to those in UCo$_{1-x}$Fe$_{x}$Ge and UTe$_{2}$ \cite{Huang2016,Ran2019}. In the present study the function of $\Phi$ is unknown, and we therefore fit the data using a polynomial to determine the goodness of scaling. 
For values of $\beta/\nu z$ between 0.45 and 0.9 and $\beta\delta/\nu z$ between 1.05 and 1.5 with a step size of 0.05, the determination of exponents is through the smallest root mean square deviation $\chi_{rms,M}$. The minimal $\chi_{rms,M}$ occurs along the diagonal where the ratio of $\beta/\nu z$ and $\beta\delta/\nu z$, i.e., $\delta$, is $\sim$ 2, as shown in Fig.~\ref{Fig1}(d). This result alone, however, could not decide other critical exponents. We hence resort to hyperscaling analysis of specific heat that allows decisive determination of $d/z$ and $\beta\delta/\nu z$. 

\begin{figure}
\includegraphics[width=0.9\columnwidth]{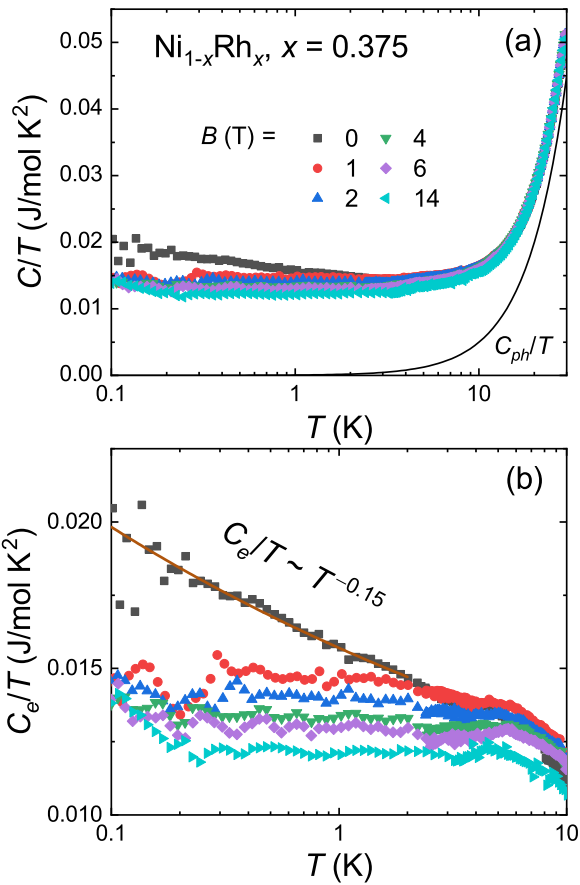}
\caption{(a) Temperature dependence of specific heat in different magnetic fields of Ni$_{1-x}$Rh$_{x}$ with $x = 0.375$, plotted as $C/T$ vs. $T$ in semi-logarithmic scale. The solid line represents the phonon contribution to the specific heat $C_{ph}/T$. (b) The electronic contribution to the specific heat $C_{e}/T$. The solid line represents a fit of $\gamma_{0} + aT^{d/z-1}$ between 0.1 and 2~K. See text for detail.}  
\label{Fig2} 
\end{figure} 

 Figure~\ref{Fig2}(a) shows the temperature dependence of the total specific heat coefficient $C/T$ in different fields. Upon cooling, zero field $C/T$ first decreases down to $\sim$ 10~K and then increases slightly toward low $T$, manifesting quantum fluctuations close to a FM QCP. As the field increases, the tendency towards FL behavior $C/T$ = constant as $T \rightarrow 0$ is gradually restored. Such a recovery of FL behavior has been observed in many antiferromagnetic and FM QCPs as the system is tuned away from the QCP and toward a magnetically disordered phase \cite{Loehneysen2007,Manuel2016}. 

The total specific heat $C$ can be expressed by $C = C_{e} + C_{ph}$, where $C_{e}$ and $C_{ph}$ are electronic and phonon contributions to the specific heat, respectively. For $x < x_{c}$ in Ni$_{1-x}$Rh$_{x}$, the split of the itinerant conduction $d$ electron bands causes FM order based on Stoner's model \cite{Santiago2017}. When $x \rightarrow x_{c}$, quantum fluctuations emerge and destroy the magnetic order. Both effects (sometimes intertwined) contribute to $C_{e}$ at low $T$. In order to properly extract $C_{ph}$, we focus on a relatively high $T$ region where the above mentioned effects are negligible, and fit the $C/T$ data between $T = 15$ and 30~K with a sum of a constant ($\gamma_{0}$, Sommerfeld coefficient) and an integral Debye term for $C_{ph}$. The obtained $C_{ph}$ is shown as a solid line in Fig.~\ref{Fig2}(a). After subtracting $C_{ph}$ from $C$, temperature dependence of $C_{e}/T$ with increasing $B$ is shown in Fig.~\ref{Fig2}(b). The data scatter below 0.4 K due to the fact that $C_{e}/T$ only amounts to $\sim 15-20$ mJ/mol K$^{2}$, comparable with the value of exchange-enhanced systems without quantum fluctuations \cite{Tari2003}, and the gradual decrease of $C_{e}/T$ from $B = 0$ to 14~T is less than 20\%, close to the resolution limit of the calorimeter. 

\begin{figure}
\includegraphics[width=\columnwidth]{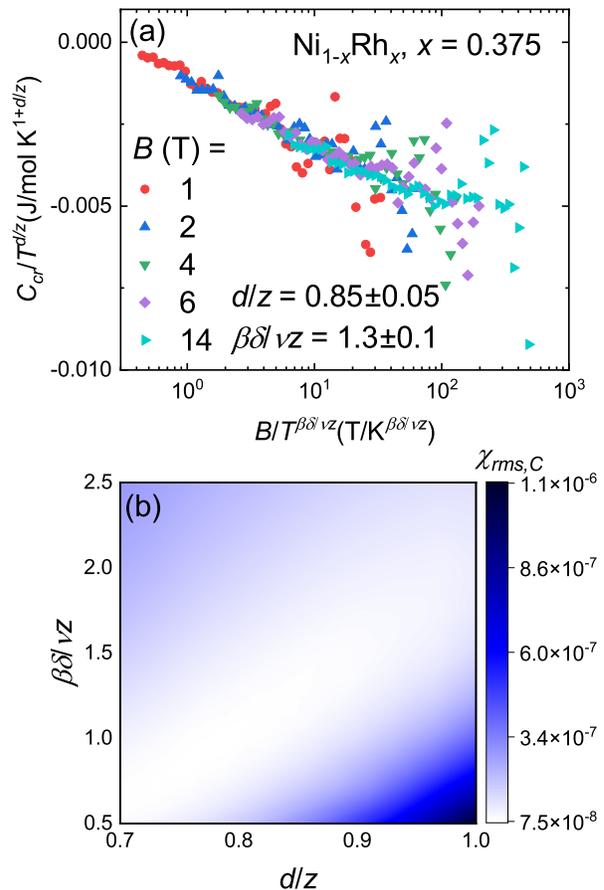}
\caption{(a) Scaling of $C_{cr}$ as a function of $T$ and $B$ of Ni$_{1-x}$Rh$_{x}$ with $x = 0.375$. (b) $\chi_{rms,C}$ as a function of critical exponents $d/z$ and $\beta\delta/\nu z$.}  
\label{Fig3} 
\end{figure} 
The hyperscaling relation of the quantum critical part of the specific heat is given by
\begin{equation}
    C_{cr}(T,B)=-T\frac{\partial^{2}\mathcal{F}}{\partial T^{2}}=b^{-d}C_{cr}\left( b^{z}T,b^{\beta\delta/\nu}B\right ).
    \label{C_{cr}1}
\end{equation}
We choose $b^{z}T = T_{0}$ again and Eq.(\ref{C_{cr}1}) becomes
\begin{equation}
    \frac{C_{cr}(T,B)}{T^{d/z}}=\Psi\left (\frac{B}{T^{\beta\delta/\nu z}} \right ).
    \label{C_{cr}2}
\end{equation}
In order to check the $T$ and $B$ scaling of the specific heat, we plot $(C(T,B)-C(T,0))/T^{d/z}$ vs. $B/T^{\beta\delta/\nu z}$ in Fig.~\ref{Fig3}(a) to eliminate non-critical quasiparticle contribution to $C_{e}$ \cite{Huang2015}. We vary $d/z = 0.65-1$ with a step size of 0.05 and $\beta\delta/\nu z = 0.5-2.5$ with a step size of 0.1, and use a polynomial fit to determine the combination of critical exponents which allow the data to collapse onto a universal curve. A local  minimum in $\chi_{rms,C}$($d/z$,$\beta\delta/\nu z$) confining $\beta\delta/\nu z$ to be 1.0-1.5 with $d/z = 0.75-0.9$ is found (Fig.~\ref{Fig3}(b)). 

Finally, Eq.~\ref{C_{cr}2} hints that the zero-field specific heat $C_{cr}(T,0) = T^{d/z}\Psi(0)$ which allows an unequivocal determination of $d/z$. Figure~\ref{Fig2}(b) demonstrates the data between 2 and 0.1~K can be well described by $C_{e}/T = \gamma_{0} + aT^{d/z-1}$ with $\gamma_{0} = 5.7$ mJ/mol K$^{2}$, $a = 10$ mJ/mol K$^{2.15}$, and $d/z =~0.85$. With this last piece of evidence, we are able to scale $C_{cr}(T,B)$. Apart from the low $T$ scattered data due to small $C_{cr}$, scaling works over three orders of magnitude in $B/T^{\beta\delta/\nu z}$ with  $\beta\delta/\nu z$ = 1.3$\pm$0.1 (Fig.~\ref{Fig3}(a)). 

\section{Discussions}
From magnetization scaling analysis we have derived $\delta = 2.0\pm0.2$. This value is small compared with the mean-field Hertz-Millis-Moriya theory either for clean or disordered systems ($\delta = 3$) which is usually enhanced via critical fluctuations \cite{Hertz1976,Millis1993,Moriya1985,Huang2015}. The dynamical critical exponent $z = 3.5\pm0.2$ can be deduced from $d/z = 0.85\pm0.05$ if we assume $d = 3$. Such an assumption is deemed to be reasonable as the crystal structure of Ni$_{1-x}$Rh$_{x}$ is face-centered-cubic.  

\begin{table}
   \caption{Comparison between experiments and theory}
\label{Table1}
    \centering
   \begin{tabularx}{\linewidth} {|l||X|X|X|X|}
   \hline
 & current result & Hertz's fixed point in the dirty limit & BKV's theory & BKV's theory in the preasymptotic region \\
   \hline
$\delta$& 2.0$\pm0.2$    & 3        & 1.5                   & 1.8\\
$z$& 3.53$\pm0.2$   & 4       & 3                     & 3.6 \\
\hline
      \end{tabularx}
\end{table}

Our experimental results are summarized and compared with theoretical predictions in TABLE~\ref{Table1}. The obtained $\delta$ and $z$ values are not comparable with the Hertz's model even in the dirty limit \cite{Loehneysen2007}. It is well known that Hertz's fixed point is unstable against the existence of dangerous irrelevant variables. The soft mode, where the correlation functions diverge as the frequency and the wave number go to zero, couples to FM order parameter fluctuations, causing multiple dynamical critical exponents and hence avoiding formation of the FM QCP \cite{Manuel2016}. Upon utilizing an appropriate amount of chemical disorder, BKV suggested one could restore the FM QCP and the critical exponents $\delta = \frac{d}{2}=1.5$ and $z = d = 3$ \cite{Kirkpatrick2014,Kirkpatrick2015}. When a system is asymptotically close to the FM QCP, both critical exponents are modified by an effective exponent $\lambda$  that depends on the distance from a QCP. BKV predicted that in $d = 3$, $\lambda = 2/3$ in a large region, and hence $\delta_{asym} = \frac{d+\lambda}{2} = 1.8$ and $z_{asym} = d+\lambda = 3.6$. Our results agree well with the BKV theory in the preasymptotic region, which implies that $x = 0.375$ locates very close to the FM QCP of Ni$_{1-x}$Rh$_{x}$ in which quantum fluctuations lead to divergence (our previous work Ref.~\cite{Huang2020}) and hyperscaling behavior of thermodynamic properties.

The key concept in the BKV theory to reach a FM QCP is via appropriate amount of chemical disorder. If the disorder effect is too strong, the QCP is avoided and spin glassiness may appear near the boundary of the quantum phase transition between FM and PM states \cite{Kirkpatrick2014,Kirkpatrick2015,Vojta2010}. Experimentally, however, it is difficult to gauge the degree of disorder from different sources. For example, how do we judge that the amount of disorder in Ni$_{1-x}$Pd$_{x}$ and Ni$_{1-x}$V$_{x}$ is larger than that in Ni$_{1-x}$Rh$_{x}$? Conventional crystal structure analysis tools, such as x-ray diffraction, electron probe microanalyzer, and scanning electron microscopy, cannot give a clue to this question. One may turn to utilize the probe of magnetic properties with elemental selectivity, e.g., x-ray magnetic circular dichroism (XMCD), to study the magnetic homogeneity. Usually, when a FM element is alloyed with a nonmagnetic metal $A$, its PM effective moment $\mu_{eff}$ derived from a Curie-Weiss fit at high $T$ decreases as the concentration of $A$ increases. Ni$_{1-x}$Rh$_{x}$ is unique among Ni$_{1-x}A_{x}$ alloys as it is the only system in which $\mu_{eff}$ remains almost constant as $x \rightarrow x_{c}$ \cite{Boelling1968}. This reflects a fact that magnetism in 4$d$ Rh atoms is largely enhanced when Rh is surrounded by Ni atoms \cite{Krishnamurthy1994}. XMCD measurements on Ni$_{1-x}A_{x}$ alloys close to $x_{c}$ may reveal the magnetic behavior in the local environment either on the Ni or $A$ site, and tell the difference of the homogeneity of magnetic properties among different systems. This testing method, if works, will further justify the applicability of the BKV theory and serve as a useful guidance when exploring new FM QCPs.      


We thank Dr. M.-K. Lee and C.-C. Yang at PPMS-16T and SQUID VSM Labs, Instrumentation Center, National Cheng Kung University (NCKU) for technical support.  We thank D. Belitz and T. R. Kirkpatrick for useful discussions. CLH would like to thank E. Morosan for her kind support in the earlier stage of this work. This work is supported by the Ministry of Science and Technology in Taiwan (grant number MOST 109-2112-M-006-026-MY3 and MOST 110-2124-M-006-009) and the Higher Education Sprout Project, Ministry of Education to the Headquarters of University Advancement at NCKU.


\end{document}